# Dispositif de supervision pour les tuteurs impliqués dans un apprentissage à la gestion de projets

**Christine Michel***

*\* Laboratoire LIESP*
*Université de Lyon, INSA-Lyon,*
*Bâtiment Léonard de Vinci 21, avenue Jean Capelle*
*F-69621, France*
*Christine.Michel@insa-lyon.fr*

RÉSUMÉ. *Cet article présente les résultats de l'observation, selon des méthodes de KM, d'une formation à la gestion de projet. L'organisation pédagogique repose sur une structure riche et complexe de tuteurs pouvant jouer des rôles variés. L'observation fine de cette formation selon des techniques de KM nous a permis de formaliser les différents rôles que jouaient les tuteurs ainsi que les problèmes de réalisation. Une proposition d'amélioration est faite sous la forme d'un dispositif de supervision instrumenté utilisable non seulement par les tuteurs mais aussi par les apprenants. Ce dispositif s'apparente à un outil métacognitif. Alors que l'apprentissage par projet, en face-à-face ou hybride, survalorise souvent l'articulation conceptualisation-expérimentation, l'originalité de notre approche est d'intégrer de manière équilibrée l'articulation réflexion-action.*

MOTS-CLÉS : *Tutorat, gestion de l'activité, apprentissage par projet, apprentissage à la gestion de projet*

**1. Introduction : La formation a la gestion de projet**

La formation en gestion de projet devient incontournable dans l'enseignement supérieur, particulièrement dans les écoles d'ingénieur et les grandes écoles. En effet, l'étude [THOMAS & MENGEL 08] sur l'évolution de cette discipline dans l'enseignement supérieur montre qu'entre 2004 et 2007 le nombre de programmes concernés est passé de 6982 à 12500 (soit +79%). La formation en gestion de projet n'a en revanche pas significativement changée durant cette période en dépits des recommandations et suggestion du PMI (Project Management Institute). L'étude de Thomas et Mengel montre que l'apprentissage à la gestion de projet doit prendre en compte à la fois des compétences « dures » et des compétences « soft ». Les compétences dures correspondent à savoir faire le planning du projet (Diagramme de Gantt), organiser la gestion du projet, prévoir les structures d'arrêt et les points d'organisation, gérer les ressources, faire le contrôle, le suivi et la clôture (recette) du projet, utiliser des outils d'automatisation et une technologie mature, comprendre et formaliser le besoin du client, organiser les reporting du projet et apprendre de ses erreurs ou bonnes pratiques [MANZIL-E-MAQSOOD & JAVED 07]. Les compétences « soft » correspondent à des compétences sociales (collaboration, communication), émotionnelles (empathie, considération de l'autre, humour, éthique), et d'organisation (leardership, vision politique, etc.) [THOMAS & MENGEL 08 ; BERGGREN & SÖDERLUND 08].

La pédagogie la mieux adaptée à cette formation est l'apprentissage par projet [BREDILLET 08]. Les connaissances et compétences s'y construisent dans l'interaction sociale et l'expérimentation directe. Elle substitue, à la pédagogie classique, une dynamique de co-développement, de responsabilité collective et de coopération [HUBER 05]. L'apprenant y est un acteur/auteur principal. De son activité découle un enrichissement significatif pour lui-même et tous les autres apprenants. Une conséquence de cet apprentissage est de segmenter la classe en sous-groupes projets encadrés par un ou des tuteurs. Mais lorsque chaque groupe fonctionne en autonomie, sur des domaines ou dans des environnements réels (entreprises par exemple) variés et lorsqu'en plus le projet se déroule sur de longues périodes, comme c'est le cas dans notre contexte d'étude, la coordination et l'harmonisation de l'activité des tuteurs sont extrêmement difficiles à réaliser. De plus, ces contextes rendent difficiles la perception de l'activité individuelle et du groupe, en particulier lorsqu'aucun support technique d'information et de communication n'est utilisé. Notre propos, au travers de l'observation et l'évaluation d'une formation à la gestion de projet riche et complexe, est de formaliser le type de pédagogie qui s'y exerce et ainsi d'identifier les spécifications d'une plateforme de supervision pour les tuteurs.

**2. Analyse d'une formation**

*2.1. Déroulement de la formation*

La formation que nous allons évaluer repose sur une organisation sociale complexe faisant intervenir des groupes projet d'élèves, des clients industriels, deux tuteurs par groupe projet, deux responsables qui coordonnent l'activité des tuteurs, un enseignant et un responsable pédagogique qui coordonne l'avancée des apprentissages de tous les groupes. Imaginée par le fondateur du département Génie Industriel Patrick Prévôt [PREVOT 08] et élément structurant de la quatrième année du cycle d'ingénieur GI INSA-Lyon, la formation à la gestion de projet de déroule sur 6 mois et correspond à un investissement d'environ 3000 heures de travail pour les élèves par projet. Les principaux objectifs pédagogiques sont : (1) apprendre à conduire un projet (étudier un marché, analyser un cahier des charges, négocier, gérer des ressources, planifier, spécifier, prototyper, etc.) ; (2) apprendre à conduire collectivement un projet (organisation collective du groupe, communication, animation, coordination dans le groupe), (3) apprendre à collecter, extraire, structurer et formaliser la connaissance, (4) développer la créativité, (5) aborder une entreprise sous tous ses aspects (histoire, organisation, stratégie, implantation, etc.) [DPT GI 08]. Nous remarquons que les objectifs 1 et 3 visent à l'acquisition de connaissances « dures » alors que les 2 et 4 d'autres de type « soft ».

La formation est composée d'un cours sur les principes et méthodes de gestion de projet et d'une mise en pratique dans un projet (appelé PCo pour « projet collectif ») réalisé par groupes de 8 élèves. Les étudiants sont en effet formés à la manipulation d'outils de management mais aussi à la pratique effective du management. Fin septembre, les 12 groupes se voient attribuer des sujets différents de PCo proposés par des industriels (ils seront les clients) pour répondre à des besoins réels. Chaque groupe travaille en autonomie dans une salle projet dédiée. Il est aidé par deux tuteurs qui lui sont exclusivement dédiés et qui encadrent respectivement l'activité de management et l'activité technique. Deux responsables, technique et management, coordonnent l'activité de leurs tuteurs respectifs. Un responsable pédagogique, que l'on peut qualifier de méta-tuteur, coordonne l'ensemble du processus. Le déroulé de la formation est structuré en quatre phases [PERRIER 08] :

- mi novembre, les équipes doivent rendre la *réponse à l'appel d'offre (RAO)* qui formalise les besoins des clients et les premières spécifications de réponses envisagées. Ce document est vu comme un document contractuel car signé par le client.
- Fin décembre ils doivent rendre le *plan directeur* qui détermine les moyens et l'organisation du projet ainsi que les documents de contrôle de l'avancé du projet (*tableau de bord*) et de la qualité des livrables (*règles de recette*). Cette première phase de la formation se clôt par une

- présentation théâtralisée devant les tuteurs (appelé « point à mi-parcours »).
- De janvier à mars les étudiants sont en *production* et répondent effectivement au besoin du client en réalisant un produit ou une étude. Ils effectuent la recette de leur réalisation fin mars.
- Jusqu'à mi-avril ils effectuent les corrections finales et rédigent les *livrables de clôture du projet* : un *rapport technique* qui décrit précisément le produit et un *rapport management* qui est une analyse du déroulement et des problèmes du projet ainsi qu'un retour individuel par élèves de ses impressions et apprentissage. Le projet se clôt par une *présentation théâtralisée* devant tous les acteurs du projet (client compris) et l'ensemble de la formation GI (en particulier les autres étudiants qui n'ont pas suivi la formation). Les élèves y présentent les résultats du projet, leurs apprentissages, les problèmes rencontrés et les moyens de les résoudre sous la forme d'une représentation costumée. Le ton, souvent humoristique, et le déguisement leur permettent d'exprimer plus facilement et librement leur point de vue, tant sur le plan des ressentiments et remerciements envers l'équipe enseignante que des bénéfices qu'ils tirent de cette formation.

*2.2. L'activité des tuteurs*

Le tuteur management intervient uniquement en présence et propose des formations adaptées au contexte et problématiques spécifiques du projet. Plus précisément, si nous reprenons la typologie des rôles des tuteurs proposée par Garrot [GARROT 08], nous pouvons dire que le **tuteur management** joue un rôle de *catalyseur social et intellectuel* car il se place plutôt en position de poser des questions ou ouvrir les discussions. Il intervient aussi comme formateur sur les questions de travail en équipe. Dans ce cadre il soutient et stimule la *coordination du groupe de travail* en jouant parfois un rôle de *médiateur*, parfois un rôle de *coach relationnel*. Ce rôle est assez systématiquement proposé pour soutenir le chef de projet mais aussi pour tout élève en difficulté relationnelle ou individuelle dans le groupe. Enfin, dans la mesure ou la formation est longue et les compétences souvent tacites (par exemple le leadership, la gestion de la relation client, la négociation….), le tuteur management joue un rôle de *méta-catalyseur*, *d'individualisateur* ou *d'autonomisateur*. Il joue aussi les rôles de *pédagogue*, *expert du contenu*, *évaluateur* et *qualimetreur* concernant les concepts et documents de synthèse de management. Le **tuteur technique** intervient pour soutenir les apprentissages relatifs aux méthodes de management de projet, mais aussi ceux nécessaires à la réalisation opérationnelle du projet (c'est-à-dire relatif au sujet/problème à résoudre dans le projet). Il assure un rôle principal de *pédagogue*, *expert du contenu* et *évaluateur* mais assure aussi des fonctions de *qualimetreur* concernant les livrables techniques. Il intervient soit en face-à-face soit à distance de manière asynchrone. Il est sollicité généralement par mail pour toutes les corrections sur les livrables. Enfin, le **responsable pédagogique** contrôle globalement l'activité de tous les groupes. Il joue exclusivement le rôle *d'évaluateur* et de *qualimetreur*. Ayant une vision transversale sur tous les groupes projet, il définit et contrôle les spécifications des livrables et le déroulé global de l'activité. Il intervient exclusivement en présence.

### *2.3. Mode de supervision*

Aucun dispositif de supervision n'est actuellement proposé aux tuteurs pour le suivi des activités ou leur notation. L'appréciation de l'activité des élèves se fait de manière non formelle, selon le nombre et la qualité des interactions qu'ils ont avec les tuteurs lors des séances en face à face. Certain tuteurs s'appuient aussi sur le nombre d'échanges par mail ou les appréciations des chefs de projet transmises de manière informelle lors de discussions. En termes de communication et coordination, chaque tuteur travaille individuellement avec son groupe et ne communique pas systématiquement avec son alter-ego technique ou management pour avoir une vision complète de l'activité du groupe. Les échanges se font principalement lors des *présentations théâtralisées*. Les tuteurs techniques et management y ont la possibilité de discuter à propos de leur groupe et d'échanger sur les autres pratiques. Ils se retrouvent, après la dernière soutenance, en huis clôt pour un jury final où ils discutent de leurs perceptions des comportements et investissements de chaque élèves, harmonisent les notations des groupes et attribuent des points de bonus ou malus (de +2 à -2) aux élèves s'étant caractérisés. Une *journée de débriefing pédagogique* est organisée pour les tuteurs management. Ils peuvent y exposer le déroulé de leur gestion de groupe, demander de l'aide ou des avis pour résoudre des conflits spécifiques et se coordonner sur les modalités d'évaluation et le mode d'accompagnement des élèves. Le responsable des tuteurs techniques et ses tuteurs se coordonnent quant à eux de *manière informelle*, selon des communications interpersonnelles. Nous pouvons observer que les rôles des tuteurs sont assez variés, que leur nombre est grand (24+2) et que chaque projet est unique et non reproductible, ce qui rend l'organisation de cette formation relativement complexe et difficile à mettre en œuvre. Dans un souci d'amélioration continue, nous tentons d'imaginer des solutions instrumentées pour faciliter la gestion et envisageons en particulier de soutenir la supervision des tuteurs. De manière à bien identifier le fonctionnement et les besoins des différents acteurs nous avons fait une observation de terrain et une analyse selon des techniques de Knowledge Management (KM).

## 3. Observation de terrain : les tuteurs

### *3.1. Méthode d'observation*

**La méthodologie** utilisée est adaptée de MKSM et KADS [DIENG et al 05]. Ces méthodes, à partir de documents produits par une organisation et d'entretiens d'acteurs, modélisent les systèmes complexes industriels. Chaque concept

(produit/acteur/activité/règles/contraintes) a été défini dans une fiche. Les fiches ICARE (Illustration Contrainte Activité Règle Entité) décrivent précisément tout objet intervenant dans un processus à modéliser. Les *fiches* RISE (Reuse, Improve and Share Experience) renseignent les problèmes ayant pu intervenir lors d'un processus et précisent les contextes, solutions proposées ou recommandations. Les éléments décrits dans les fiches ICARE et RISE ont été globalement organisés dans des *diagrammes* qui montrent leurs inter-relations. L'adaptation des méthodes MKSM et KADS à notre contexte a été réalisée avec l'aide de l'ancien directeur du service KM d'Airbus (Toulouse), René Peltier. **L'observation effective** a été réalisée par des élèves de 5eme année GI dans le cadre d'un TP du cours KM. Ils ont utilisé comme sources d'observation *la documentation formelle produite dans le cadre du PCo* (rapport management des équipes, guide de formations des tuteurs, règles d'évaluation), ainsi que *des retours d'expériences* (REX) et *transfert d'expertise* (EXTRA) des acteurs du projet. Les REX ont été réalisés sur les élèves eux-mêmes (ils ont fait un retour sur leur propre expérience de l'année précédente lorsqu'ils étaient en PCo) ainsi que sur des tuteurs techniques et management. Les EXTRA ont été réalisés avec les responsables techniques et management et avaient pour fonction de formaliser l'ensemble de leurs activités et responsabilités. Le mode de recueil des REX des élèves a consisté en une formalisation directe de leur expérience dans les fiches ICARE et RISE. Le recueil a porté sur une soixantaine d'élèves (sur 2 ans) ayant fait le PCo. Celui des tuteurs (REX-tuteur) et responsables (EXTRA) s'est fait par des entretiens semi-directifs qui ont été ensuite exploités par les élèves du TP KM pour remplir des fiches. Ce recueil a concerné les responsables des tuteurs, 6 tuteurs techniques et 3 étudiants chef de projet actuellement en train de suivre la formation PCo.

### 3.2. Analyse des problèmes liés aux tuteurs

Les problèmes vécus et exprimés ont été décrits dans 36 fiches RISE. La plupart ne concernaient pas directement les tuteurs mais plutôt la gestion du travail d'équipe au sein du groupe et l'organisation pédagogique du projet. Celles qui concernaient l'activité des tuteurs ont mentionné des problèmes d'évaluation, de présence, de cohérence et de coordination. Les étudiants ont en effet exprimé un sentiment d'injustice concernant *l'évaluation individuelle* dans la mesure ou tous les membres du projet, même les étudiants qui se sont peu ou moins investis, profitent de la note du groupe (à + ou – 2 point selon leur investissement). Les tuteurs ont aussi parallèlement exprimés leur impuissance à concrètement pouvoir évaluer individuellement les étudiants, impuissance liée au caractère intuitif et tacite de l'appréciation, au manque de traçabilité des actions des élèves, et à l'absence de discussion avec leurs collègues. Certains élèves ont souligné le *manque de communication* ou le manque de *présence* de certains tuteurs. D'autres ont mentionné un *manque de cohérence* et des défauts dans la *coordination* et la *diffusion d'information* concernant par exemple les consignes (ambigües ou contradictoires) données aux différents groupes ou la mise en pratique des concepts théoriques. Deux groupes d'élèves ont plus spécifiquement étudié comment résoudre les problèmes identifiés dans les fiches RISE en instrumentant l'activité : le premier [FENG *et al.* 09] a évalué la recevabilité d'un outil de **gestion du travail collaboratif** comme Sharepoint, l'autre [BILLOIS *et al.* 09] les évolutions que l'on pouvait faire sur **les tableaux de bords** qui sont des dispositifs de supervision construits par chaque chef de projet pour voir l'état d'avancement de leur projet et évaluer l'activité de leur équipe. Les indicateurs présentés montrent en effet le temps de travail de chaque membre, la répartition des taches, les retards éventuels par rapport au planning mais aussi le moral de l'équipe et les compétences acquises (ou à acquérir) pour faire la réalisation. Nous présentons brièvement quelques résultats de la première étude et plus longuement la deuxième.

*3.3. Étude de solutions instrumentées*

L'étude de **SharePoint** a montré que ce type d'application pouvait *a priori* améliorer le processus (et en tout cas directement intervenir sur la moitié des problèmes identifiés dans les fiches RISE) [FENG et al 09]. L'entretien semi-directif avec les tuteurs a mis en évidence le fait que ce type d'application pouvait être particulièrement utile pour favoriser le fonctionnement collaboratif et distribué (géographiquement) du projet en particulier concernant la gestion des documents et leur processus de validation par un workflow. Une autre idée forte a été la possibilité de création d'un espace d'échange accessible aux membres en tout point. Enfin, les tuteurs ont apprécié le fait que le support informatique permettrait de garder une trace des actions de chaque élève et aiderait à l'évaluation individuelle. L'étude de l'utilisation du **tableau de bord** (TDB) [BILLOIS *et al.* 09] a montré que cet outil est en fait peu utilisé, il reste un exercice théorique, réalisé exclusivement par le chef de projet et presque jamais consulté par les autres membres du projet. Le groupe qui a travaillé sur l'amélioration du TDB a soumis une proposition d'évolution aux tuteurs et aux actuels chefs de projet. Elle est représentée dans la figure 1 suivante. L'ancien TDB est représenté dans la zone « *indicateur* » et se présente sous une forme graphique. L'évolution concerne sur la gauche de la figure l'ouverture de zone de saisie pour les informations factuelles (moral, heures de travail, …), l'ajout d'indicateurs, l'ajout d'une nouvelle interface de visualisation de synthèse individuelle (*page perso*). Au centre on observe l'ouverture de zones d'expression (bloc note/blog) personnelles et collectives. A droite, il est proposé aussi de faire apparaitre les agendas partagés des tuteurs transverses à plusieurs projets pour faciliter la prise de rendez-vous et le contact. Le mode de consultation (lecture en trait pointillé, écriture en trait plein) est représenté par les flèches.

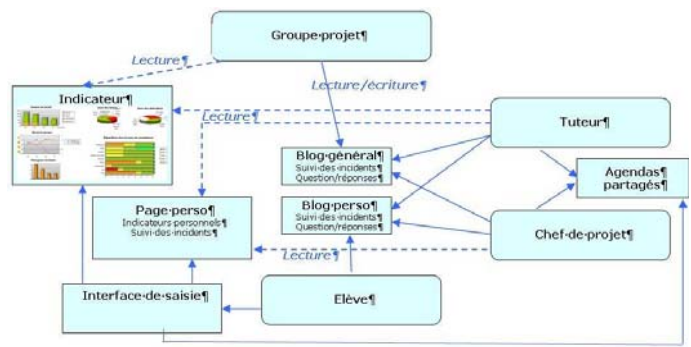

**Figure 1.** *Proposition d'évolution du tableau de bord*

Concernant le **mode de fonctionnement et l'utilisation de l'outil,** les tuteurs interrogés approuvent l'idée d'ouverture en termes de mise à jour et consultation du TDB. Certains indiquent en effet que le TDB est peu utilisé parce qu'il n'est tout simplement pas identifié et localisé du fait d'un manque de communication du chef de projet. De plus, ils pensent que ce mode de renseignement/diffusion favorise les apprentissages sur cet outil de tous les membres et soutient la communication dans le groupe et entre tuteur-élèves. Néanmoins, ils mentionnent que toute information doit être préalablement validée par le chef de projet avant publication. L'articulation du TDB avec des outils de communication (chat et blog) ou des outils de gestion du travail collaboratif (calendrier, versionning de document, espace collaboratif) leur semble fondamentale même s'ils n'ont pas de proposition formelle d'utilisation. Les espaces d'expression du projet et de chaque membre sont particulièrement appréciés tant par les tuteurs que les chefs de projet. En effet, ces derniers donnent l'exemple de l'indicateur « incident » ou « moral » qui doit pouvoir être contextualité de manière à permettre au tuteur et au chef de projet d'avoir conscience et de comprendre les raisons des dysfonctionnements et d'initier un dialogue avec les membres du projet. Réciproquement cela permet aux membres d'avoir conscience de l'état d'esprit du chef de projet et ainsi d'être plus rapidement responsabilisé en cas de problème ou valorisé en cas de réussite. En revanche un espace de communication inter-projet ne semble pas souhaitable aux élèves alors qu'il est jugé comme nécessaire pour les tuteurs.

Concernant la **conception du TDB**, le fait qu'il soit réalisé par le groupe projet est jugée discutable, dans la mesure où la forme au final est relativement imposée, et pourrait être remplacé par un exercice critique. Les acteurs sont globalement satisfait du type et de la forme des indicateurs actuels mais propose des modification/ajout. Les étudiants souhaitent par exemple avoir une appréciation de l'acquisition effective des « compétences pour la réalisation du projet », faite par le tuteur et le chef de projet. Plus globalement, ils sont demandeurs de plus de formalisme sur ce point et en particulier ils expriment un besoin d'accompagnement dans une réflexion, au démarrage du projet, sur l'ensemble des compétences nécessaires et des moyens de formation. Cela leur permettrait d'être rassurés, d'avoir une vision globale et de mieux répartir les apprentissages et les taches pour chaque membre du projet. Ils expriment néanmoins clairement leur besoin d'une solution « simple » et refuse en particulier qu'elle puisse prendre la forme d'un portfolio. Sur un autre plan, les étudiants sont favorables à l'ajout d'un indicateur « professionnalisme » (respect des délais, des chartes d'équipe, anticipation et alertes auprès du chef de projet de problèmes rencontrés ou retards dans les tâches) qui qualifierait le comportement de l'étudiant sur d'autres critères que les réalisations techniques.

## 4. Discussion et conclusion : vers un nouveau mode de supervision

Les observations de fonctionnement montrent des attentes au niveau des tuteurs concernant une rationalisation de la coordination et de la collaboration (avec les élèves et aussi entre pairs), mais aussi concernant la réalisation effective de leur activité et de leur rôle. Les solutions envisagées par une instrumentation combinant les fonctionnalités de SharePoint et des actuels tableaux de bords semblent prometteuses. Concernant SharePoint, nous nous appuierons sur les études de [GEORGE 01] et [FOLLET & PEYRELONG 09] pour définir le meilleur soutien instrumenté à la production et collaboration des élèves et des tuteurs via des outils de coordination et de communication comme les agendas partagés, les chat ou les forums par exemple. Cette instrumentation permet de plus d'accompagner efficacement plusieurs des rôles des tuteurs, en particulier les rôles de pédagogue, évaluateur et qualimetreur en traçant les activités des élèves et en ayant un accès direct aux contributions comme le mentionne [GARROT 08]. L'utilisation des tableaux de bord « évolués » est particulièrement adaptée à notre modèle pédagogique qui est basé sur le cycle de Kolb [CORTEZ *et al*. 08] en quatre phases : expérience personnelle, observation réflexive, conceptualisation, expérimentation. L'articulation conceptualisation-expérimentation est très marquée comme dans tous les apprentissages par projet. L'originalité de notre approche est aussi de bien prendre en compte l'articulation entre action (expérimentation ou conceptualisation) et réflexion sur soi. En effet, comme Berggren [BERGGREN & SÖDERLUND 08] nous pensons que les formations à la gestion de projet survalorisent le rôle de l'expérimentation et qu'il faut rééquilibrer l'organisation avec des actions réflexives pour accompagner une **évolution de comportement** en termes de management, communication et collaboration (ce que nous avons appelé compétences « soft »). Plus que des techniciens expérimentés, nous visons à rendre nos élèves capables d'« apprendre à apprendre et à évoluer » selon les contextes en favorisant leur capacité à avoir une analyse critique de leurs actions selon les situations auxquelles ils sont confrontés. Cette capacité résulte principalement de l'accompagnement des tuteurs. Nous pensons affiner la proposition sur les tableaux de bord pour en faire un **outil métacognitif** qui interviendrait précisément pour soutenir cet accompagnement. Selon Azevedo [AZEVEDO 07], l'outil métacognitif accompagne les processus d'apprentissages complexes en présentant des informations sur l'apprenant relatives : à sa *cognition* (activation des savoirs antérieurs, planning, construction de sous-objectifs, stratégies d'apprentissage), à sa *métacognition* (jugement de son apprentissage), à sa *motivation* (efficacité, besoin, intérêt, efforts) et à son *comportement*. Plusieurs de ces notions apparaissent déjà dans le tableau de bord et demandent à être affinées comme la motivation ou la cognition. D'autres, comme le fai d'avoir des jugements sur les apprentissages ou de collaborer dans la définition de stratégies d'apprentissage, mis en évidence par l'observation comme des besoins, tant pour les tuteurs que pour les élèves. Enfin, ce type d'outil peut directement aider à la réalisation du rôle de méta-catalyseur, d'individualisateur et d'autonomisateur du tuteur management. En effet, dans la mesure où il est impossible actuellement d'instrumenter la finesse de perception (basées sur la forme sociale du groupe et la complexité des psychologies humaines) nécessaire à l'interaction du tuteur, nous recommandons que ce rôle et l'activité pédagogique de débriefing qui lui est liée continue à se dérouler lors de séance en face-à-face. Néanmoins le tuteur doit pouvoir être aidé dans sa perception individuelle et sa mémorisation par des supports comme les contenus des zones d'expression personnelles et collectives.

**Bibliographie**